\pdfoutput=1
\documentclass[conference]{IEEEtran}

\usepackage[sort&compress, numbers]{natbib}
\usepackage{verbatim}
\usepackage{caption}
\usepackage{subcaption}
\usepackage[]{algorithm2e}
\usepackage{graphicx}
\usepackage{booktabs}
\usepackage{relsize}
\usepackage{colortbl}
\usepackage{enumerate}
\usepackage{enumitem}
\usepackage{amsmath}
\usepackage{tikz}
\usepackage{pgf}
\usepackage{pbox}
\usepackage{rotating}
\usepackage{balance}
\usepackage{tablefootnote}
\usepackage[bookmarks=false]{hyperref}
\usepackage{flushend}
\usepackage{tcolorbox}
\usepackage{tabularx}
\usepackage{hanging}
\usepackage{multirow}
\usepackage{xcolor}
\usepackage{cleveref}
\usepackage{arydshln}
\usepackage{amssymb}
\crefformat{footnote}{#2\footnotemark[#1]#3}

\usepackage{dblfloatfix}
\usepackage[T1]{fontenc}
\usepackage{soul}

\hypersetup{draft} 

\let\oldFootnote\footnote
\newcommand\nextToken\relax

\renewcommand\footnote[1]{%
    \oldFootnote{#1}\futurelet\nextToken\isFootnote}

\newcommand\isFootnote{%
    \ifx\footnote\nextToken\textsuperscript{,}\fi}

\newcommand{\rectangle}{{%
  \ooalign{$\sqsubset\mkern3mu$\cr$\mkern3mu\sqsupset$\cr}%
}}

\newcommand{\rqi}{To what degree is \di{} important for effort estimation?}
\newcommand{\rqiii}{Is a change of \di{} more likely to influence re-estimation than \ios{}?}
\newcommand{\rqiv}{What types of \di{} are useful for effort estimation?}

\newcommand{\infochg}{information change}

\newcommand{\sigpvalue}{$p<0.05$}
\newcommand{\nonsigpvalue}{$p\geq0.05$}

\newcommand{\peopleCount}{148}

\newcommand{\responseRate}{10\%}
\newcommand{\eligibleCount}{121}
\newcommand{\useDocumentCount}{117}
\newcommand{\countryCount}{25}

\newcommand{\sectionOne}{Informed consent}
\newcommand{\sectionTwo}{Preliminary questions}

\newcommand{\sectionFour}{Importance of \di{}}
\newcommand{\sectionFive}{Re-estimation}
\newcommand{\sectionSix}{Useful information and its quality}
\newcommand{\sectionSeven}{Demographics}

\newcommand{\di}{documented information}

\newcommand{\ios}{the information from other sources}
\newcommand{\iosnothe}{information from other sources}

\newcommand{\revised}[2]{#2}

\newcommand{\revisedTextOnly}[1]{{#1}}

\newcommand{\always}{`Always (\~{}100\%)'}

\newcommand{\difr}{functional requirements}
\newcommand{\dius}{user stories}
\newcommand{\didd}{definition of done}
\newcommand{\diui}{UI wireframes}
\newcommand{\diat}{acceptance criteria}
\newcommand{\ditd}{\revisedTextOnly{task dependencies}}

\newcommand{\qchange}{changing}
\newcommand{\qmiss}{missing}
\newcommand{\qoutdate}{outdated}
\newcommand{\qwrong}{wrong}

\newcommand{\frdefinition}{\revisedTextOnly{the task information about the functionality of the system}}

\begin{document}

    
    \title{\huge{Towards Just-Enough Documentation for Agile Effort Estimation:
    What Information Should Be Documented?}}


    \author{\IEEEauthorblockN{Jirat Pasuksmit, Patanamon Thongtanunam, Shanika Karunasekera}
    \IEEEauthorblockA{The University of Melbourne, Australia\\
    jpasuksmit@student.unimelb.edu.au, patanamon.t@unimelb.edu.au, karus@unimelb.edu.au}
    }


    \maketitle

\begin{abstract}
Effort estimation is an integral part of activities planning in Agile iterative development.
An Agile team estimates the effort of a task based on the available information which is usually conveyed through documentation.
However, as documentation has a lower priority in Agile, little is known about how documentation effort can be optimized while achieving accurate estimation.
Hence, to help practitioners achieve just-enough documentation for effort estimation, we investigated the different types of \di{} that practitioners considered useful for effort estimation.
We conducted a survey study with \eligibleCount{} Agile practitioners across \countryCount{} countries.
Our survey results showed that (1) despite the lower priority of documentation in Agile practices, 98\% of the respondents considered \di{} moderately to extremely important when estimating effort, (2) 73\% of them reported that they would re-estimate a task when the \di{} was changed, and (3) \difr{}, \dius{}, \didd{}, \diui{}, \diat{}, and \ditd{} were ranked as the most useful types of \di{} for effort estimation. Nevertheless, many respondents reported that these useful types of \di{} were occasionally \qchange{} or \qmiss{}.
Based on our study results, we provide recommendations for agile practitioners on how effort estimation can be improved by focusing on just-enough documentation.

\end{abstract}


\section{Introduction}

Effort estimation is an important practice for planning activities in Agile iterative development.
It helps a software development team to approximates the size of a task, enabling an optimal workload allocation for a short iteration of Agile~\cite{Cohn2006, fucci2018needs}.
The estimated effort of the tasks in previous iterations also can  be used to calculate the delivery capability, which helps the team to better plan the next iteration~\cite{coelho2012effort, rubin2012essential}.
Hence, accurate effort estimation benefits the team in creating a reliable iteration plan to deliver the product increment~\cite{Cohn2006, rubin2012essential}.

The effort of a task is typically estimated based on the available information~\cite{rubin2012essential, usman2018effort}.
One of the common approaches to manage and convey information is to use documentation~\cite{andriyani2017understanding, gralha2018evolution, klotins2019progression}.
For example, backlog items are used to document the information related to a feature (e.g., \dius{}, \diat{})~\cite{kasauli2017requirements, sedano2019product}.
A lower-level backlog item (like development tasks) may also include the detail information related to the implementation of the feature (e.g., a description of a function to be developed)~\cite{lauesen2002software}.

On the other hand, in Agile practices, documentation has a lower priority than working software~\cite{fowler2001agile}.
Based on this concept, it is possible that the information that is recorded in documentation (\textbf{\textit{\di{}}}, henceforth) may not be well maintained, resulting in inadequate information for the team to understand development tasks~\cite{kasauli2017requirements}.
Several studies also discovered that \di{} in Agile projects can be incorrect, incomplete, outdated, and changing~\cite{aghajani2019software, kasauli2017requirements, zhi2015cost}.
Yet, little is known about how documentation effort can be minimized while achieving accurate estimation.

In this paper, we aim to understand how just-enough documentation can be provided (i.e., what information should be documented). 
Hence, practitioners can save documentation effort while achieving accurate effort estimation.
Therefore, we conducted a survey study with Agile practitioners to investigate the importance and the types of \di{} that are useful for effort estimation in the current practices.
Our findings will be beneficial to Agile practitioners, tools developers, and researchers.
Agile practitioners can pay attention to the quality of useful documented information to achieve more accurate estimation.
Tool developers and researchers can focus their attention on developing tools and techniques to help practitioners maintain documentation for effort estimation.
Based on the responses of \eligibleCount{} Agile practitioners, we answered the following research questions:

\noindent \textbf{(RQ1) \rqi{}}\\
\indent \underline{Motivation}: Accurate effort estimation requires an adequate understanding of a task~\cite{jorgensen2009impact,usman2018effort,svensson2019bam, Usman2016}.
While documentation can convey the information to understand the task~\cite{hoda2012documentation, andriyani2017understanding, gralha2018evolution}, it has a lower priority than delivering working software in the Agile concept~\cite{fowler2001agile}.
Hence, we investigated how Agile practitioners perceive the importance of the \di{} for effort estimation. \linebreak
\indent \underline{Results}: 97\% of the respondents used documentation to assist effort estimation, and 98\% of them reported that \di{} was moderately to extremely important when estimating effort. 
The majority of them attempted to keep \di{} and documented effort up to date most of the time.
Although documentation has a low priority in Agile, our result shows that the \di{} is considered important for effort estimation. 

\noindent \textbf{(RQ2) \rqiii{}}\\
\indent \underline{Motivation}:
Prior studies reported that effort may be re-estimated when the information is changed~\cite{hoda2016multilevel, Bick2018}.
However, in Agile, the information can be conveyed through various channels, e.g., documentation, in-person, discussion threads~\cite{andriyani2017understanding, gralha2018evolution, tizard2019can}.
\revisedTextOnly{Hence, we set out to investigate how often a task was re-estimated when the \di{} was changed in comparison to \ios{}.}
\linebreak
\indent \underline{Results}: 
The reported frequency of re-estimation when \textit{\di{}} was changed is statistically higher than the reported frequency of re-estimation when \textit{\iosnothe{}} was changed.
In general, the respondents noted that re-estimation will be performed if the \infochg{} is significant or is from a reliable source like documentation.
This result shows that when \di{} is changed, a task is often re-estimated.

\noindent \textbf{(RQ3) \rqiv{}}\\
\indent \underline{Motivation}: Comprehensive documentation can be considered as an overhead in Agile~\cite{fowler2001agile, hoda2012documentation}.
On the other hand, inadequate documentation may lead to a misunderstanding of a task~\cite{svensson2019bam, kasauli2017requirements}.
Hence, just-enough and just-in-time documentation would be suitable for Agile practices~\cite{heck2017framework}.
\revised{R3.1}{Yet, little is known what types of information are considered useful for effort estimation and should be documented.
}
\linebreak
\indent \underline{Results}: Functional requirements (i.e., \frdefinition{}), \dius{}, \didd{}, \diui{}, \diat{}, and \ditd{} were ranked as the six most useful \di{} for effort estimation. 
Yet, many respondents reported that these useful types of \di{} were occasionally (or more) \qchange{} or \qmiss{} during (or after) effort estimation.
Our results suggest that the information that is useful for effort estimation (especially for the six useful types of \di{}) should be documented and reviewed prior to finalizing effort estimates.

Our findings highlight that \di{} is important for effort estimation and a change of \di{} can lead to re-estimation. 
Aligning with the Agile practices, our recommendations can be done in a just-in-time manner when the information is needed for effort estimation.
In other words, attention should be paid to ensure that the useful \di{} for effort estimation has sufficient detail prior to finalizing effort estimates to avoid making unnecessary assumptions.
In addition, to help practitioners achieve accurate effort estimation, our work highlights the need for tools and techniques to timely assess the quality of \di{}.

\textbf{Paper organization.}
Section~\ref{sec:definition} defines the terms used in this paper.
Section~\ref{sec:researchmethod} presents our research methodology. Section~\ref{sec:result} presents the study results.
Section~\ref{sec:discussion} provides a broader implication of the results.
Section~\ref{sec:relatedwork} discusses related work.
Section~\ref{sec:threats} discloses the threats to validity.
Finally, Section~\ref{sec:conclusions} draws the conclusion.

\section{Terms \& Definitions}
\label{sec:definition}
This section defines the terms that are used in this paper.
\begin{description}
    \item[Information] refers to a description that provides an understanding of a software product under development, e.g., software features~\cite{ebert2008effectively, IEEETerm}.
    \item[Documented information] refers to the information related to a software product that is captured and recorded in documentation (i.e., a collection of documents)~\cite{IEEETerm, forward2002relevance}.
    The \di{} can be in various forms.
    For example, the information related to a feature (e.g., \dius{}, \diat{}) can be documented in a backlog item~\cite{sedano2019product, kasauli2017requirements}.
    A description of a function to be developed can be documented in a lower-level backlog item like a development task~\cite{lauesen2002software}.
    The documentation we consider can be recorded on a physical paper or in a task management system (e.g., JIRA)~\cite{hoda2012documentation,hoda2016multilevel,khalil2019exploring,schon2017agile, gralha2018evolution}.

\end{description}

We do \textit{not} refer the documentation to textual descriptions related to source code (e.g., code comments)~\cite{ebert2019confusion, zhi2015cost}, software development process~\cite{ebert2008effectively}, user manuals, and informal communication (e.g., discussion thread, chat)~\cite{tizard2019can, gralha2018evolution}.

    \section{Research Method}

\label{sec:researchmethod}

In this section, we describe our survey design, survey questions, recruitment process, and data analysis.
\newcommand{\ssymbol}[1]{^{\@fnsymbol{#1}}}
\begin{table*}
\caption{Survey questions (excluding demographics questions\cref{drive}).}
\label{tab:questions}
\normalsize
\vspace*{-0.05in}
\hspace*{-0.20in}
\centering
\resizebox{1\textwidth}{!}{
\centering
\begin{tabular}{l|l}
\hline
Item & Question \\ \hline

& \textbf{S2: \sectionTwo{}} \\
PQ1*
& Does your team perform effort estimation in any stage of the software development lifecycle? (yes/no) \\

PQ2
& Have you ever participated in effort estimation of your team? (yes/no) \\
\hline

& \textbf{S4: \sectionFour{} (RQ1)} \\
Q1
& Does your team use any documentation when performing effort estimation? (yes/no) \\

Q1.1
& What is the source of information (other than the information in the documentation) that your team uses in effort estimation? (Open-ended) \\
Q2$\dagger$
& In general, how important is the task-related documented information when you estimate the task? \\
Q3$\ddagger$
& How often do you update the documentation when new information is available? \\
Q4$\ddagger$
& How often do you update the estimated effort in the documentation when the task is re-estimated? \\
\hline

\hline

& \textbf{S5: \sectionFive{} (RQ2)} \\
Q5*$\ddagger$
& \underline{Before you work on a task}, do you re-estimate the task if \underline{the documented information} that is relevant to the task is changed or updated? \\
Q6*$\ddagger$
& \underline{Before you work on a task}, do you re-estimate the task if \underline{the information from other sources than documentation} is changed or updated? \\
Q7*$\ddagger$
& \underline{While you are working on a task}, do you re-estimate the task if \underline{the documented information} that is relevant to the task is changed or updated? \\
Q8*$\ddagger$
& \underline{While you are working on a task}, do you re-estimate the task if \underline{the information from other sources than documentation} is changed or updated? \\

\hline

& \textbf{S6: \sectionSix{} (RQ3)} \\
Q9
& On average, which of the following documented information do you find most useful in effort estimation? (Ranking) \\
Q10
&
For each of the selected types of \di{}, please select the following quality issues that you occasionally (or more) encounter \\
& during (or  after) effort estimation for an average task that you estimate.\\
& Multiple selections of quality issues: \rectangle{} missing, \rectangle{} incorrect, \rectangle{} outdated, \rectangle{} changing, or \textcircled{} generally no issue \\

\hline

\multicolumn{2}{l}{* Questions with asterisk included an optional open-ended answer for further elaboration.} \\

\multicolumn{2}{l}{$\dagger$ A Likert scale of importance: \textcircled{} Extremely important, \textcircled{} Very important, \textcircled{} Moderately important, \textcircled{} Slightly important, \textcircled{} Not at all important} \\

\multicolumn{2}{l}{$\ddagger$ A Likert scale of frequency: \textcircled{} Always (\~{}100\%), \textcircled{} Most of the time (\~{}75\%), \textcircled{} About half the time (\~{}50\%), \textcircled{} Sometimes (\~{}25\%), \textcircled{} Never (\~{}0\%)} \\

\end{tabular}
}
\end{table*}

\subsection{Survey Design}
\label{sec:surveydesign}
The survey was designed based on the recommended survey design practice in software engineering to avoid known problems~\cite{ghazi2018survey, kitchenham2002preliminary}.
Our survey consisted of seven sections which included 11 questions to address our research questions (see Table~\ref{tab:questions}) and 11 questions to collect demographics about the respondents.
Participation was voluntary and the estimated time to complete a survey was 10-15 minutes.
The survey responses were recorded anonymously. 
The respondent could drop out anytime, and the incomplete response was automatically deleted.
The survey was created using an online survey platform named Qualtrics. 
To ensure the overall quality of responses, we used the features of Qualtric to prevent bots and multiple/duplicate responses from a single respondent.\footnote{\url{http://qualtrics.com/support/survey-platform/survey-module/survey-checker/response-quality}}

\underline{Pilot Tests:} Comprehensibility of the survey questions was our main construct concern when designing the survey~\cite{kasunic2005designing}.
As suggested by Ghazi et al.~\cite{ghazi2018survey}, we conducted two pilot tests with five software engineers from our connections.
These five participants of the pilot tests were software engineers who used an Agile method and used effort estimation (i.e., the target population of our survey). 
They had 5 to 12 years (7.6 years on average) of experience. 
After the pilot tests, the comprehensibility of the survey was improved by adding details, rewording, adding and merging options, and providing definitions.
Finally, our survey and related documents were reviewed and approved by the Human Ethics Advisory Group of the authors' institution.

\subsection{Survey Questions}
Table~\ref{tab:questions} provides an overview of our survey questions that address our three RQs.\footnote{The full survey and the replication package is available at \url{http://figshare.com/s/74c2509bdb23de953da3}\label{drive}}
The questions with an asterisk (*) included an optional open-ended question to collect an elaboration for an answer of the corresponding question.
We now briefly describe each section of our survey.

\textbf{S1: \sectionOne{}.} This section informed the respondent about the researchers' contact, the purpose of this study, how a response will be recorded, and how long will it take to complete the survey.
Before proceeding to the next section, the respondent was asked to consent to participate in the study.

\textbf{S2: \sectionTwo{}.} We asked two preliminary questions to check whether the respondent had been involved in effort estimation.
To be specific, we asked whether the team of the respondent performed effort estimation at any stage of the software development process (PQ1) and whether the respondent had ever participated in effort estimation of the team (PQ2).
Given that a respondent who answered `\texttt{no}' to one of these two questions is unlikely to have background knowledge about effort estimation of the team, the rest of the survey questions were skipped and the respondent was directed to the \sectionSeven{} section.

\textbf{S3: Terms and Definitions.}
This section provided the terms and definitions that we used in this survey (see section~\ref{sec:definition}).
This is to ensure that the respondent will have the correct understanding of the survey questions.
We also informed the respondent that the survey questions focused on effort estimation at the \textit{lowest} granularity of a work item that s/he estimated.
Since the respondent may be involved in multiple projects, we asked the respondent to answer the survey questions based on the project in which s/he was most involved in.
Finally, we asked the respondent to confirm that s/he understood the terms and definitions.

\textbf{S4: \sectionFour{}.} The survey questions in this section aimed to address our RQ1.
We asked whether the respondent used any documentation when performing effort estimation (Q1).
If the respondent answered `\texttt{no}', we asked what is the other sources of information that the respondent used (Q1.1).
\revisedTextOnly{Then, the respondent was directed to the \sectionSeven{} section (S7)}.
If the respondent answered `\texttt{yes}' to Q1, we asked the respondent to rate the importance of \di{} for effort estimation (Q2) using a 5-points Likert scale of importance (see Table~\ref{tab:questions}).
Since the respondent who answered `\texttt{yes}' to Q1 used documentation for effort estimation, we further asked the respondent about documentation maintenance.
Specifically, we asked how often the respondent updated the document when new information is available (Q3) and how often s/he updated the documented effort when a task is re-estimated (Q4).
To answer Q3 and Q4, we provided a 5-points Likert scale of frequency (see Table~\ref{tab:questions}).

\textbf{S5: \sectionFive{}.} The survey questions in this section aimed to address our RQ2.
We asked how often the respondent re-estimates a task when the \di{} was changed in comparison to \ios{}~\cite{andriyani2017understanding, gralha2018evolution, tizard2019can}.
We also considered whether the information was changed before or after the work had started as this may be a factor that influences a decision of re-estimation~\cite{hoda2016multilevel,wang2014role,paetsch2003requirements}.

\begin{figure}[t!]
    \centering
    \includegraphics[width=0.8\linewidth]{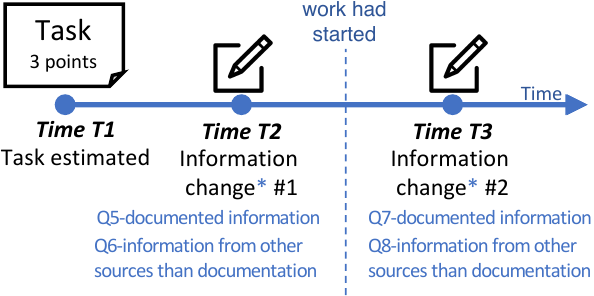}
    \caption{Illustrative Scenario for \sectionFive{} section (S5).}
    \label{fig:scenario}
\end{figure}

Hence, we formulated four survey questions (Q5-Q8), where Q5 and Q7 focused on the changes of \di{}, while Q6 and Q8 focused on the changes of information from other sources.
Furthermore, Q5-6 focused on the re-estimation when the information was changed before the work had started, while Q7-8 focused on the re-estimation when the information was changed after the work had started.
To ease the respondent's understanding, we provided an illustrative scenario (Figure~\ref{fig:scenario}) in our survey. 
Additionally, we included open-ended questions for the respondents to elaborate their answers to Q5-8.

\textbf{S6: \sectionSix{}.} The survey questions in this section aimed to address our RQ3. 
We asked the respondent to select five types of \di{} and rank them based on the usefulness for effort estimation at the lowest granularity that s/he estimated (Q9).
We provided nine common types of \di{} as the options.
We also provided open-ended options where the respondent can fill out other types of \di{} that were not on the list.
We derived the nine common types of \di{} from the literature which are as follows:
\begin{itemize}
    \item User Stories~\cite{rubin2012essential,james2010scrum, sedano2019product, behutiye2017non},
    \item Definition of Done~\cite{rubin2012essential,james2010scrum, behutiye2017non},
    \item Acceptance criteria (including acceptance tests and test cases)~\cite{rubin2012essential,james2010scrum, Davies2014, sedano2019product, gralha2018evolution},
    \item Functional requirements~\cite{lauesen2002software},
    \item Non-functional requirements~\cite{lauesen2002software, behutiye2017non},
    \item Task dependencies~\cite{sedano2019product, gralha2018evolution, tanveer2016understanding},
    \item Code examples~\cite{Davies2014, Zimmermann2010, aghajani2019software},
    \item UI wireframes (or pre-designed UI)~\cite{sedano2019product, heck2017framework}, and
    \item API reference guides~\cite{aghajani2019software}.
\end{itemize}
In our survey, we provided a definition for each of these types to ensure that the respondent had the same understanding.\cref{drive}
Note that, since we focused on Agile practices, \difr{} are referred to \frdefinition{}~\cite{lauesen2002software} as opposed to software requirement specification (SRS) that is prepared upfront in non-Agile software development practices.
Both user stories and \difr{} capture the information about the system behaviors.
The key difference between the two types is that a \dius{} is a lightweight expression that focuses on the desired business value.
In contrast, \difr{} focus on the technical aspect of a functionality to be developed.

For each selected type of \di{}, we further asked the respondent to select the quality issues that s/he occasionally (or more) encountered during (or after) effort estimation (Q10).
In this work, we study four common quality issues of \di{}: (1) \qchange{}, (2) \qmiss{}, (3) \qoutdate{}, and (4) \qwrong{}~\cite{aghajani2019software,kasauli2017requirements, zhi2015cost}.
In our survey, we also provided an option of `generally no issue' if the respondent did not encounter any of these four issues.

\textbf{S7: \sectionSeven{}.}
The survey questions in this section aimed to collect the general background and effort estimation practice of the respondent.\cref{drive}
We asked the respondent about the software development methodology, experience in the IT industry (years), experience in effort estimation (years), the duration of the most involved project (years), office location (country), working team size, and team location (co-located, distributed-inshore, or distributed-offshore).

We asked what the effort estimation technique(s) and unit(s) that the team of the respondent used, and when the team performed effort estimation (e.g., iteration planning, backlog grooming, project planning).
As the respondent's team might use different effort estimation techniques, the questions in this section were multiple selections.
We also asked an optional open-ended question about the benefit of effort estimation.

\subsection{Recruitment Process}
To recruit software engineers to participate in our survey, we used purposive sampling and snowballing methods.

\textbf{Purposive Sampling.} We used purposive sampling as our main recruitment approach.
This method allowed us to identify software engineers who are likely to use an Agile method and effort estimation~\cite{etikan2016comparison}.
In particular, we targeted software engineers who (1) work (or have worked) in an international software company or a top local software company; and (2) have at least one year of experience in software development.
We mainly considered the international and top local companies as they are established companies which are more likely to use effort estimates in feature prioritization, while start-up companies tend to use business values as the prioritization target~\cite{klotins2019progression}. 
To find the top local companies, we searched for a list of top companies for each country.

We recruited software engineers via LinkedIn (i.e., a professional networking platform) as it allows us to access the target group of interest~\cite{palomares2017requirements, ghazi2018survey}.
We identified the target group of software engineers using publicly available LinkedIn profiles.
Then, we sent a survey invitation to the targeted software engineers via the LinkedIn direct message.

To ensure the diversity of respondents, we controlled the number of invitations per software company.
More specifically, for a small company (<200 employees), we randomly selected 10-20 targeted software engineers.
For a large company (>200 employees), we randomly selected 20-50 targeted software engineers.
Since LinkedIn limits the number of invitations,\footnote{https://www.linkedin.com/help/linkedin/topics/6096/6097/4800} we sent the survey invitation to about 20-30 software engineers per day for 60 consecutive days. 
Based on this process, we sent a survey invitation to 1,500 software engineers via LinkedIn.
Broadly speaking, 36\%, 20\%, 33\%, and 10\% of direct invitations via LinkedIn were sent to software engineers in the Americas, Europe, Asia, and Oceania, respectively.

\textbf{Snowballing.} To further reach out to the target population, we performed a snowballing approach~\cite{denscombe2014good}.
\revisedTextOnly{Specifically, if the respondent informed us that s/he completed the survey, we asked them to spread the survey link to other software engineers who use Agile methods and effort estimation.
In total, we asked 16 respondents to spread the link.}

\subsection{Data Preprocessing}\label{sec:preprocess}
In this study, we focused on effort estimation in Agile projects.
However, when we sent the survey invitation, we could not determine whether a software engineer \revisedTextOnly{has worked} in an Agile project or his/her team used effort estimation.
Therefore, before analyzing the survey results, we selected the responses of software engineers of interest based on their answers to the preliminary and demographic questions.
To be specific, we selected the responses using the following criteria.

\begin{description}
    \item \textit{Criterion 1 - Use an Agile method}:
    Since we focused on Agile practices, we excluded the respondents who used non-Agile methods, i.e., Waterfall model, Incremental model, and not sure or N/A.
    \item \textit{Criterion 2 - Have Effort Estimation Experience}:
    To ensure that the respondents had experience in effort estimation, we excluded the respondents who answered `\texttt{no}' to the preliminary questions about the involvement in effort estimation (i.e., PQ1 and/or PQ2).
\end{description}

\subsection{Card Sorting}\label{cardsorting}

Similar to prior work~\cite{ebert2019confusion, vassallo2020developers}, we applied card sorting to analyze the open-ended answers of Q5-8.
As suggested by Spencer~\cite{spencer2009card}, we first conducted an open card sorting procedure to extract themes of the answers (cards).
For each open-ended question, the first two authors of this paper independently grouped the answers based on the thematic similarities and defined a theme for each group.
Then, the first two authors discussed to develop the agreed-upon themes of the answers.
To validate the agreed-upon themes, the third author of this paper performed closed-card sorting, i.e., sorting the answers into the themes that were developed by the first two authors.
Finally, we measured the inter-rater reliability using Cohen's Kappa coefficient~\cite{mchugh2012interrater}.
The overall Kappa values ranged from 0.65-0.80, indicating substantial agreements between the sorters.

    \section{Results}\label{sec:result}

In this section, we present the results of our survey study.
In particular, we first provide an overview of the survey results. 
We then analyze the survey results to answer each of our RQs.

\subsection{Survey Results Overview}
\label{sec:demographic}

\begin{figure}[!t]
    \centering
    \begin{subfigure}{\columnwidth}
        \centering
        \includegraphics[width=\textwidth]{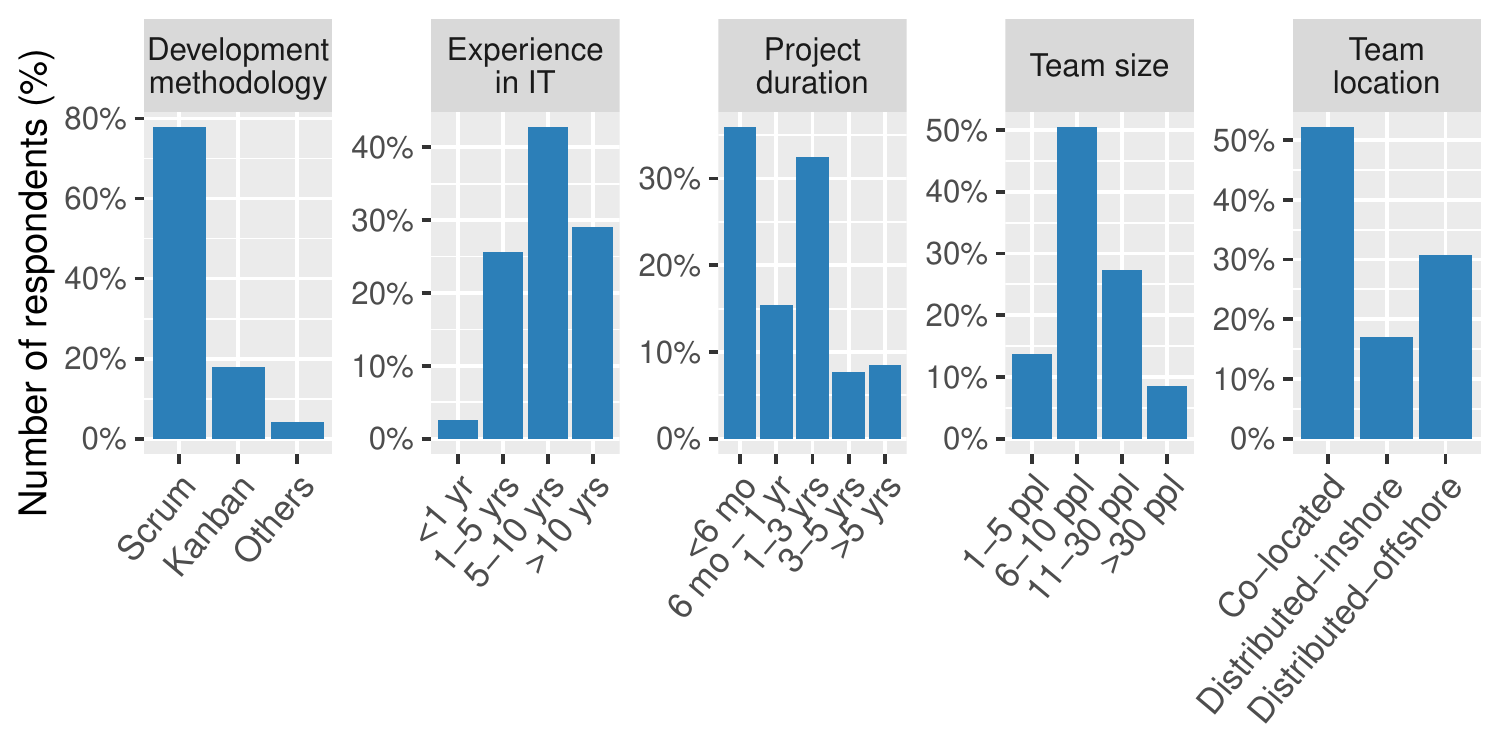}
    \end{subfigure}
    \hfill 
    \begin{subfigure}{\columnwidth}
        \centering
        \includegraphics[width=\textwidth]{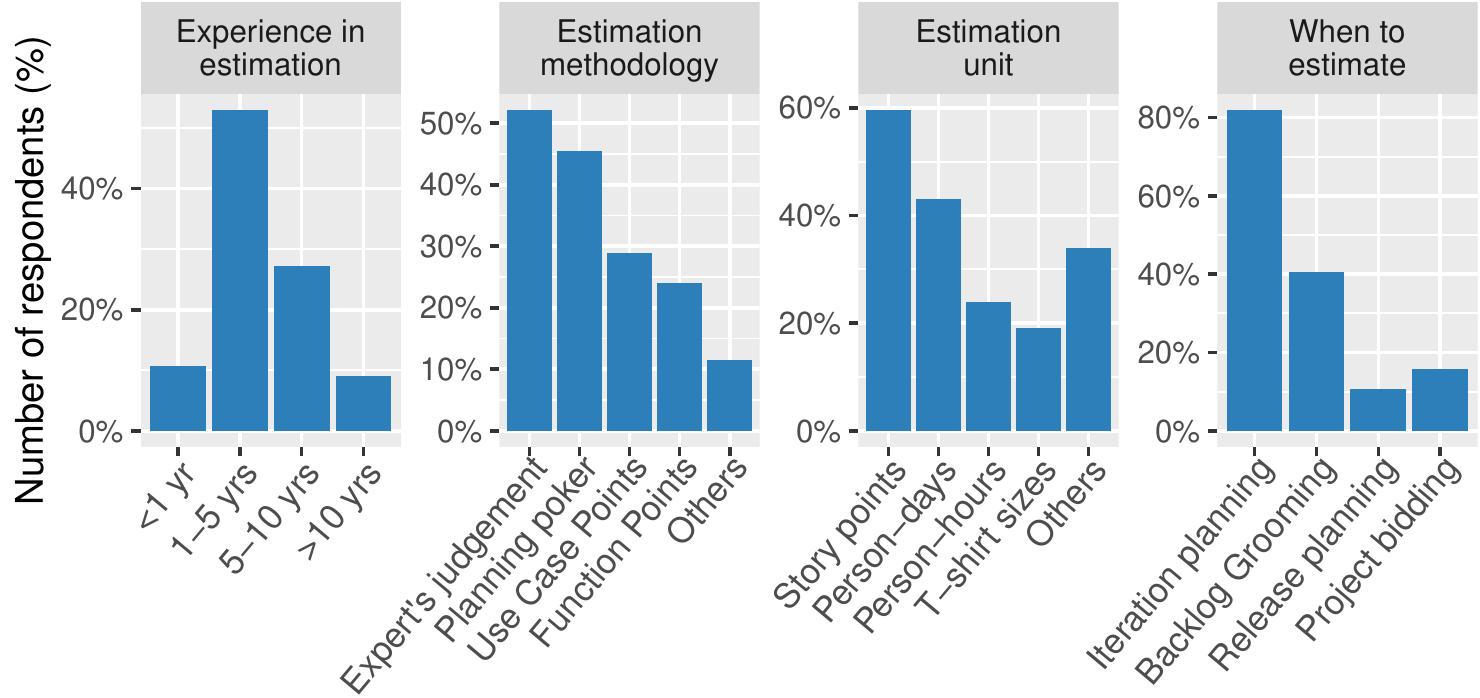}
    \end{subfigure}
    \caption{The distributions of respondents across different groups of demographics  \revisedTextOnly{and effort estimation practices}. A respondent can select multiple estimation methodologies, estimation units, and time points when estimating effort.}
    \label{fig:demographics}
\end{figure}

We received \peopleCount{} responses in total from the 1,500 direct LinkedIn invitations and snowballing (a response rate of \responseRate{}).\footnote{Since we used a snowballing approach and the responses are anonymized, we do not know the number of snowballed responses. Hence, the reported response rate may be varied.}
The median time taken by the respondents to complete the survey was 14 minutes.
After the data pre-processing (see Section~\ref{sec:preprocess}), we retained \textbf{\eligibleCount{}} respondents of interest, i.e., software engineers who used an Agile method and had been involved in effort estimation.

\textbf{Respondent Demographics.}
Figure~\ref{fig:demographics} presents the distributions of respondents across different demographic groups.
Most of the respondents used Scrum (78\%) or Kanban (18\%) as their development methodology.
A majority of them (72\%) had more than five years of experience in the IT industry.
The software project on which a respondent reported typically had a duration of three years or under (83\%).
The typical team size was 6-10 people (50\%) and 11-30 people (27\%).
About 52\% of the respondents worked in a co-located team, while others worked in a distributed team.
Our respondents were from 25 countries, e.g., Singapore (18), the United States of America (14), Thailand (13), Australia (12), India (11), others (53).

\textbf{Effort Estimation Practices.}
Figure~\ref{fig:demographics} also shows that our respondents had experience in diverse effort estimation practices.
Figure~\ref{fig:demographics} shows the effort estimation practices of our respondents.
The majority of the respondents had 1-5 years (53\%) or 5-10 years (27\%) of experience in estimating effort.
They performed effort estimation during iteration planning (83\%) and/or backlog grooming (37\%).
The common estimation methods were Expert Judgement (50\%) and/or Planning Poker (48\%).
The common effort unit(s) were story points (61\%) and/or person-days (43\%).
Based on the open-ended questions, the respondents reported that they used effort estimation to 1) know the size and complexity of development tasks, 2) achieve better resource allocation, 3) accurately estimate the delivery capability and the time to deliver, and 4) build trust with stakeholders based on an accurate and transparent workflow.

\subsection{(RQ1) \rqi}

\begin{figure}[t!]
    \centering
    \includegraphics[width=0.9\linewidth]{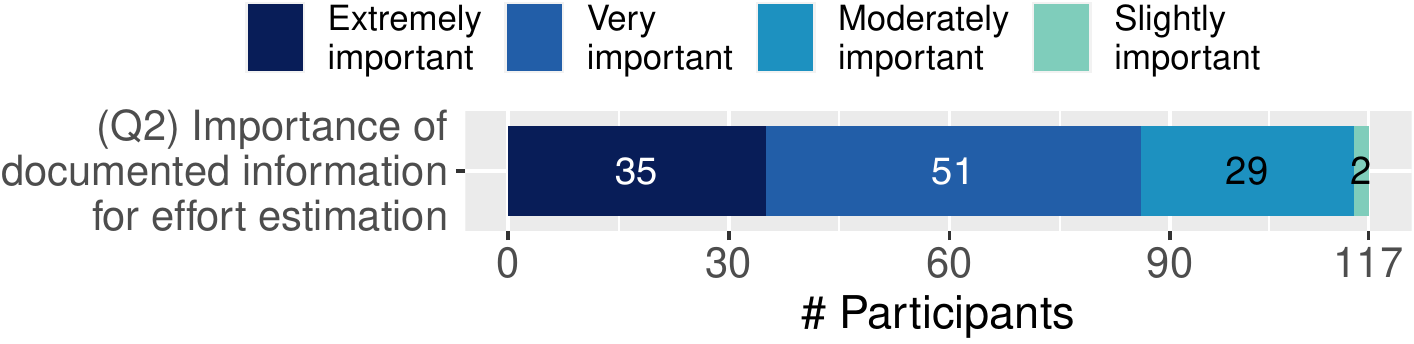}
    \caption{Importance of \di{} when performing effort estimation.}
    \label{fig:importance}
\end{figure}

A vast majority of the respondents ($\frac{\useDocumentCount{}}{\eligibleCount{}}$; 97\%) reported that they used documentation when performing effort estimation (Q1).
Figure~\ref{fig:importance} shows that most of the respondents ($\frac{115}{\useDocumentCount{}}$; 98\%) who used documentation when estimating effort rated the importance of \di{} as moderately to extremely important for effort estimation (Q2).
\revised{R1.6}{On the other hand, only two respondents rated it as slightly importance and none rated it as not at all important.}
This result indicates that practitioners considered \di{} to be moderately to extremely important for effort estimation.

Based on Q1.1, the respondents who did not use documentation when estimating effort (4 out of \eligibleCount{} respondents) reported that, instead of using documentation, they relied on 1) team velocity in the past iterations, 2) verbal communication, or 3) code-based statistics from code indexer (e.g., Google Kythe).

\begin{figure}[t!]
    \centering
    \includegraphics[width=0.9\linewidth]{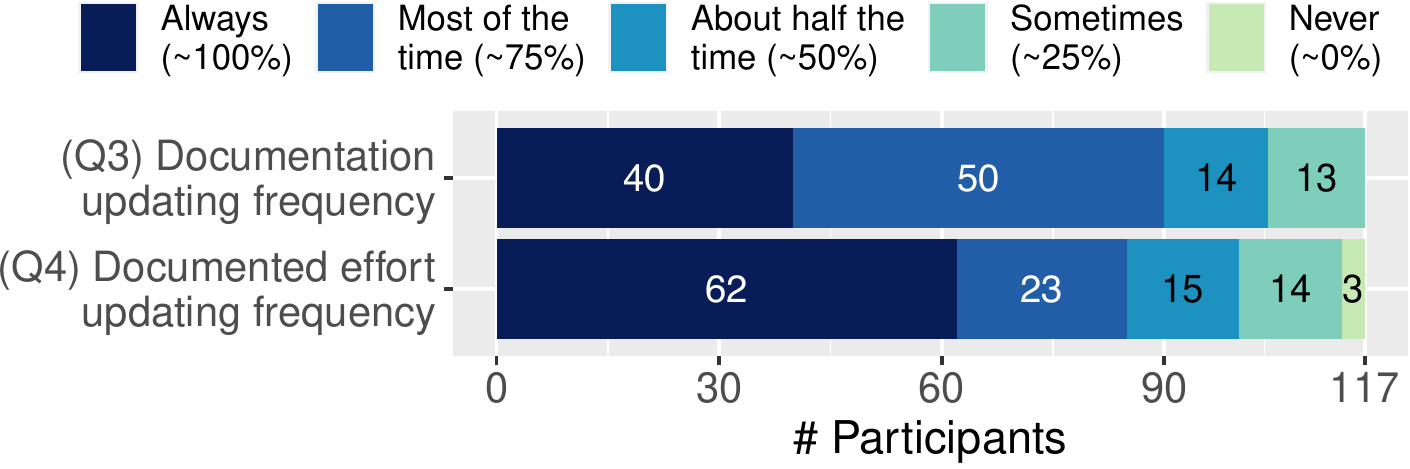}
    \caption{The frequencies that the respondents updated the documentation when new information was available and updated the documented effort after re-estimation.}
    \label{fig:docupdate}
\end{figure}

The survey results of Q3 and Q4 also show that the respondents who used documentation when estimating effort tend to keep both \di{} and documented efforts up to date.
Figure~\ref{fig:docupdate} shows that 34\% ($\frac{40}{\useDocumentCount{}}$) of the respondents always updated documentation when new information was available (Q3); and 42\% ($\frac{50}{\useDocumentCount{}}$) of them updated the document most of the time.
Similarly, 53\% ($\frac{62}{\useDocumentCount{}}$) of the respondents always updated the documented effort after re-estimation (Q4).
These results suggest that the respondents attempted to keep the documentation up to date.

Prior studies pointed out that the importance of \di{} may be varied based on the team settings~\cite{heck2018systematic,inayat2015systematic,kasauli2017requirements, gralha2018evolution}.
Hence, we further checked whether the \di{} has a different level of importance given a different team setting.
We considered three aspects of team settings, i.e., the team location, size, and project duration.
For each aspect, we used the Kruskal-Wallis rank sum test to compare the distributions of the importance level of \di{} across the groups of respondents.
For example, we examined whether the distributions of the importance level of \di{} are statistically different among the three groups of team location (i.e., co-located, distributed-inshore, and distributed-offshore).
The Kruskal-Wallis rank sum test is a non-parametric test that is robust to non-homogenous and a small sample.
Similar to prior work~\cite{harpe2015analyze, jamieson2004likert, xia2019practitioners}, we assigned the numerical values to the Likert scale, i.e., 5 (for Extremely important) to 1 (for Not at all important) when performing a statistical test.
For the three aspects of team settings, all the tests yielded \revised{R1}{not statistically significant} (\nonsigpvalue{}), indicating that the distributions of importance levels are not statistically different for the respondents from different team settings.

\begin{tcolorbox}[boxsep=1pt,left=1pt,right=1pt,top=1pt,bottom=1pt]
    \textbf{Findings:} 97\% of the respondents used documentation to assist effort estimation, and 98\% of these respondents reported that \di{} is moderately to extremely important when estimating effort.
    Furthermore, a majority of the respondents attempted to keep \di{} and documented effort up to date most of the time. 
\end{tcolorbox}


\subsection{(RQ2) \rqiii}

\begin{figure}[t!]
    \centering
    \includegraphics[width=0.9\linewidth]{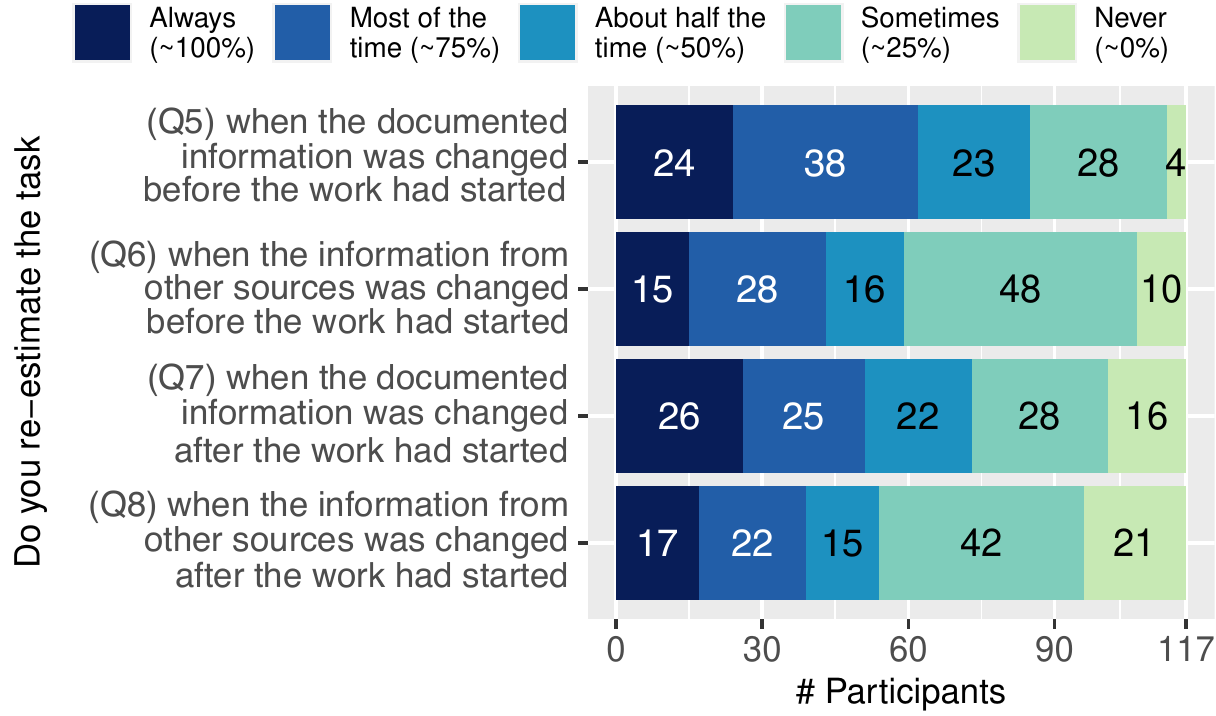}
    \caption{The frequencies that the respondents would re-estimate if the \di{} (or \ios{}) was changed before (or after) the work had started.}
    \label{fig:reestimation}
\end{figure}

We found that the respondents tend to re-estimate a task when \textit{the \di{}} was changed more often than when \textit{\ios{}} was changed.
Figure~\ref{fig:reestimation} shows that 73\% ($\frac{85}{117}$) of the respondents reported that they would re-estimate at least 50\% of the time when \textit{the \di{}} was changed before the work had started (Q5).\footnote{Note that we excluded the four respondents who did not use any documents in effort estimation since we focus on the information in documents.\label{exclusion_note}}
On the other hand, 50\% of the respondents reported that they would re-estimate at least 50\% of the time when \textit{\ios{}} was changed before the work had started (Q6).
Similarly, for the after-start-working case, 62\% of the respondents reported that they would re-estimate at least 50\% of the time when \textit{the \di{}} was changed (Q7); while 46\% of the respondents reported that they would re-estimate at least 50\% of the time when \textit{\ios{}} was changed (Q8).

To statistically confirm our finding, we used the one-sided Wilcoxon signed-rank test to examine whether a respondent reported the frequency in Q5 (i.e., when \textit{the \di{}} is changed) higher than the frequency in Q6 (i.e., when \textit{\ios{}} was changed) for the before-start-working case.
Similarly, we performed the test to compare the difference of frequencies in Q7 and Q8 for the after-start-working case.
We opted to use the one-sided Wilcoxon signed-rank test as it is a non-parametric test which is robust to non-homogenous and a small sample.
Before performing the tests, we converted the Likert scale of frequency into numerical values (e.g., \always{} to 100).
The tests showed that the reported frequency in Q5 is statistically higher than the reported frequency in Q6; and the reported frequency in Q7 is statistically higher than the reported frequency in Q8 (\sigpvalue{}).
These results confirm that the change of \textit{the \di{}} is more likely to influence re-estimation than the change of \textit{\ios{}}.

\begin{table}
\small
\caption{Themes derived from the answers of the open-ended questions of Q5-8.}
\label{tab:table3_code}
\setlength\tabcolsep{1pt}
\centering
\begin{tabular}{p{8cm}|r}
\hline
\textbf{Theme}       & \# \\ \hline
\multicolumn{2}{l}{\textit{Reasons to re-estimate}}        \\ \hline
T1) Only re-estimate if the change is significant or from a reliable source like documentation. & 52 \\
T2) Always re-estimate to correct the accuracy, velocity, or development plan.    & 26 \\

\hline

\multicolumn{2}{l}{\textit{Reasons to not re-estimate}}        \\ \hline

T3) The team has a policy (or norm) to not re-estimate after the work had started. & 12 \\

T4) Absorb the change (attempt to complete within the original time frame) and/or roll the work over to the next iteration.   & 9\\

T5) Over-estimate to cover future change.      & 2 \\ \hline

\textbf{Total responses}        & 101 \\ \hline
\end{tabular}
\end{table}

For the optional open-ended questions of Q5-8, we received 101 responses.
We sorted the responses into five themes of reasons based on the card sorting procedure (see Section~\ref{cardsorting}).
Table~\ref{tab:table3_code} lists the themes which are grouped into (1) reasons to re-estimate and (2) reasons to not re-estimate.

In general, 52 responses indicated that the decision of re-estimation was based on the significance of the \infochg{} (T1): \textit{"If the change is not significant, and does not change the scope, the same points may remain. However, if due to the change, the original estimate is not accurate, new point will be estimated. [...]"}.
In addition to the significance of the change, some responses also emphasized that the re-estimation decision also depended on the source of information: \textit{"If it’s not documented, it likely won’t be important. This is the motivation for everything to be documented."}
Another 26 responses pointed out that re-estimation was required in order to keep the delivery capability and development plan realistic (T2): \textit{"Any re-estimation is always needed to have a correct velocity sprint productivity [...]."}

The common reason for not re-estimating was that they had a policy (or norm) to not re-estimate after the work had started (T3): \textit{"At my team, estimation is frozen since task is scheduled for current sprint and sprint is ongoing."}
Another nine responses indicated that they tried to absorb the change and/or rolled the remaining work over to the next iteration (T4): \textit{"Mostly, our team just try to done the task even it's exceed the estimation. However, if the sprint ended and this task still exist and need to carry over to next sprint, we will re-estimate the remaining task [...]."}
Other two responses noted that they intentionally over-estimated to cover the future change (T5): \textit{"Because I usually expect delays based on missing or incorrect documentation, I overestimate my tasks to take into account these issues. This allows me not to have to re-estimate."}

\begin{tcolorbox}[boxsep=1pt,left=1pt,right=1pt,top=1pt,bottom=2pt]
    \textbf{Findings:} The reported frequency of re-estimation when \textit{the \di{}} was changed is statistically higher than the reported frequency of re-estimation when \textit{\ios{}} was changed.
    In general, the respondents noted that re-estimation will be performed if the \infochg{} is significant or from a reliable source like documentation.

\end{tcolorbox}

\subsection{(RQ3) \rqiv}

\begin{table}
\caption{The number of respondents who ranked the type of \di{} as one of the five most useful information; and the number of respondents who indicated that the type of \di{} occasionally (or more frequent) had a quality issue during (or after) effort estimation. }
\label{tab:diissue}
\vspace*{-0.03in}
\hspace*{0.15in}
\resizebox{0.9\linewidth}{!}{
\centering
\setlength\tabcolsep{2pt}
\begin{tabular}{l|r|r}
\hline
\multirow{2}{*}{\textbf{Type of \di{}}} & \textbf{Ranked as}  & \textbf{Have Quality}\\ 
& \textbf{Useful} & \textbf{ Issue(s)}  \\ \hline
Functional requirements        & 91 & 67 (74\%) \\
User stories                   & 74 & 41 (55\%) \\
Definition of Done             & 76 & 41 (55\%) \\
UI wireframes                  & 61 & 46 (75\%) \\
Acceptance criteria            & 79 & 65 (82\%) \\
Task dependencies              & 61 & 51 (84\%) \\
Non-functional requirements    & 40 & 36 (90\%) \\
API reference guides           & 33 & 24 (73\%) \\
Code examples                  & 6 & 3 (50\%)   \\
\hline
\end{tabular}
}
\end{table}

\begin{figure}
    \centering
    \includegraphics[width=0.9\linewidth]{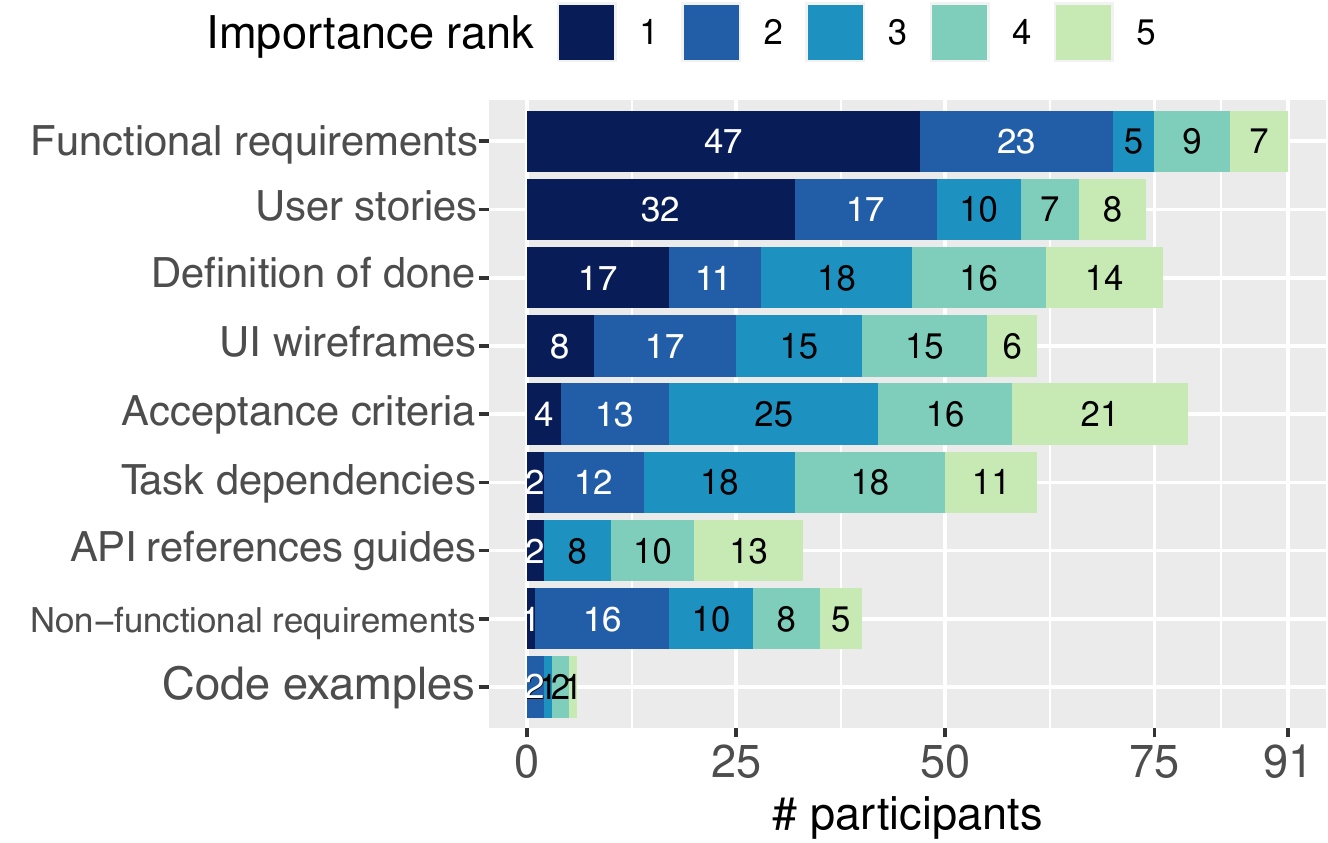}
    \caption{The distributions of importance rank of \di{}, where the rank of 1 indicates that that type of \di{} is the most useful.}
    \label{fig:diimportance}
\end{figure}

\revised{R2.1}{Table~\ref{tab:diissue} shows that \difr{}, \dius{}, \didd{}, \diui{}, \diat{}, and \ditd{}} were frequently rated as useful \di{} for effort estimation (Q9).\cref{exclusion_note}
Moreover, Figure~\ref{fig:diimportance} shows that \difr{} (i.e., the description of a function to be developed~\cite{lauesen2002software}) were ranked as the most useful type of \di{} for effort estimation where 52\% ($\frac{47}{91}$) of the respondents ranked this type as the top one.
The \dius{} was also ranked as the second most useful type where 43\% ($\frac{32}{74}$) of the respondents ranked this type as the top one.
These results suggest that the \di{} that provides an understanding of the software system's behavior is important when estimating effort.

Nine respondents also provided additional types of information that are useful for effort estimation in the open-ended options.
Most of the reported information is related to the team context, e.g., the knowledge, expertise, or workload of developers.
In addition, one respondent pointed out that steps-to-reproduce and observed behaviors are useful \di{} when estimating the effort of a bug-fixing task.

\begin{figure}
    \centering
    \includegraphics[width=1.0\linewidth]{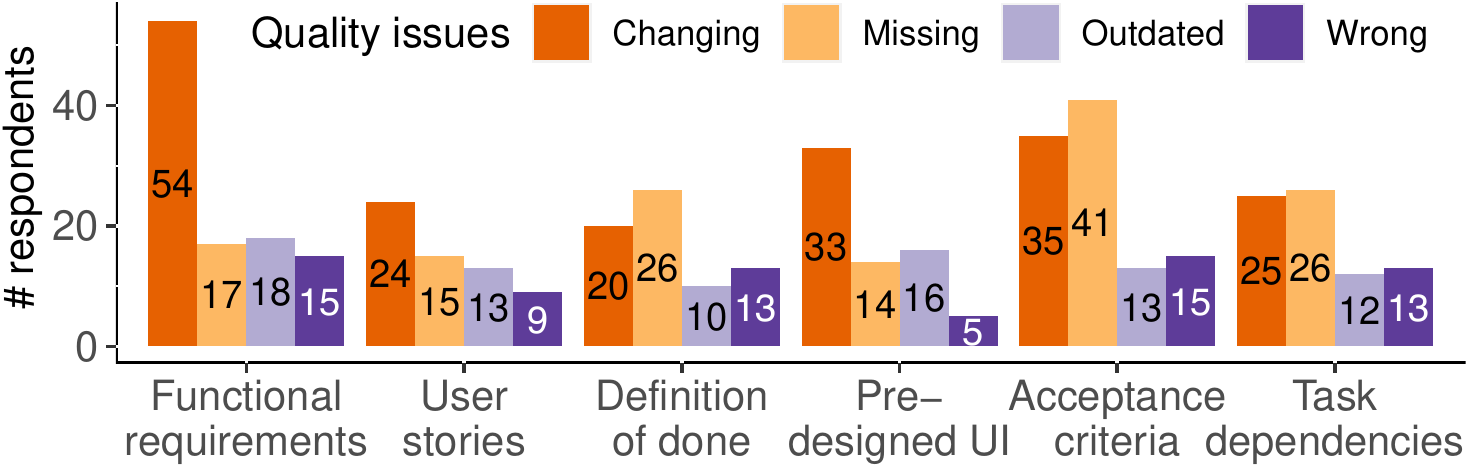}
    \caption{The number of respondents who reported that the type of \di{} occasionally (or more) had quality issues during (or after) effort estimation. Note that multiple quality issues could be selected.}
    
    \label{fig:diproblem}
\end{figure}

Table~\ref{tab:diissue} shows that many respondents reported that these useful types of \di{} occasionally (or more) had at least one kind of quality issue during (or after) effort estimation.
For example, 74\% ($\frac{67}{91}$) of the respondents who ranked \difr{} as useful indicated that this type of \di{} had quality issues.
Figure~\ref{fig:diproblem} shows that 81\% ($\frac{54}{67}$) of the respondents indicated that the \difr{} in the documentation were occasionally \qchange{}.
Furthermore, Table~\ref{tab:diissue} shows that other useful types \di{}, i.e., \dius{}, \didd{}, \diui{}, \diat{}, \revised{R2.1}{and \ditd{}} were reported as having quality issues by many respondents (55\%-84\%; see Table~\ref{tab:diissue}).
Figure~\ref{fig:diproblem} also shows that \dius{} and \diui{} were occasionally changing, while \didd{}, \diat{}, \revised{R2.1}{and \ditd{}} were occasionally \qmiss{}.
These results suggest that despite the usefulness of these types of \di{} for effort estimation, they were occasionally \qchange{} or \qmiss{}.

\begin{tcolorbox}[boxsep=1pt,left=1pt,right=1pt,top=1pt,bottom=1pt]
    \textbf{Findings:} Functional requirements, \dius{}, \didd{}, \diui{}, \diat{}, \revised{R2.1}{and \ditd{}} were ranked as the six most useful \di{} for effort estimation. Yet, many respondents reported that these useful types of \di{} were occasionally (or more) \qchange{} or \qmiss{} during (or after) effort estimation.
\end{tcolorbox}

\section{Discussion}\label{sec:discussion}

We now discuss a broader implication and provide a recommendation based on our results.

\subsection{Implications}
\textbf{Despite the lower priority of documentation in Agile, the \di{} is used and considered important for effort estimation.}
More specifically, our RQ1 shows that \useDocumentCount{} out of the \eligibleCount{} respondents used documentation for effort estimation.
Moreover, regardless of the team settings (i.e., the team location, size, or project duration), 98\% of the respondents ($\frac{115}{\useDocumentCount{}}$) considered \di{} as moderately to extremely important for effort estimation (see Figure \ref{fig:importance}).
From the open-ended question of Q6, some respondents emphasized the importance of \di{}, e.g., \textit{"I prioritize the documentation over information from other sources. The information outside documentation is needed to be filtered if it is important or it is just good to have/know. [...].}"
Furthermore, the responses of Q3-4 show that a majority of the respondents tend to maintain documentation as they reported that they updated documentation of both information and estimated effort most of the time (see Figure \ref{fig:docupdate}).
These results highlight the importance of \di{} for effort estimation for Agile practitioners.

\revised{R2.2}{\textbf{The \qchange{} \di{} can have an impact on estimation accuracy.}}
Figure \ref{fig:reestimation} shows that 73\% ($\frac{85}{117}$) of the respondents re-estimated the task at least 50\% of the time when the \di{} was changed, which is more often than the change of \ios{}.
The respondents also noted in the open-ended questions of Q5-8 that they would re-estimate if the change was significant or was from a reliable source like documentation.
Our statistical tests also confirm that the reported frequency of re-estimation when \di{} was changed is statistically higher than that of \ios{}.
These results suggest that the practitioners are more likely to re-estimate the task if \di{} was changed in comparison to a change of \ios{}.

\subsection{Recommendations}
Our survey results highlight the importance of \di{} for effort estimation.
The results also show that the quality of \di{} can have an impact on estimation.
Hence, we provide two recommendations on how documentation should be maintained while achieving accurate estimation.
Aligning with the Agile practices, our recommendations can be done in the just-enough (i.e., focusing on the useful types of \di{}) and just-in-time (e.g., before iteration planning) manner.

\textbf{Prior to finalizing effort estimates (e.g., at iteration planning), attention should be paid to ensure that the information that is useful for effort estimation is documented to avoid making unnecessary assumptions.}
Our RQ3 shows the six types of \di{} that were ranked as the most useful for effort estimation (see Figure~\ref{fig:diimportance}).
However, several respondents reported that these useful types of \di{} had quality issues, especially for the \qchange{} issue (Table~\ref{tab:diissue}).
The responses of the open-ended question of Q5 also highlight that a change of such useful information often caused re-estimation: \textit{"Depends on changes. [...] the change of \dius{}/\difr{} changes the original estimation often."}
Since the effort estimates are used in iteration planning, such uncertainty may cause the plan to become unreliable~\cite{svensson2019bam, Cohn2006}.
Hence, to facilitate effort estimation and improve its accuracy, these useful types of information should be documented and reviewed.
One possible actionable approach is to confirm the information with stakeholders.
This is also aligned with one respondent who noted that \textit{"[...] It's actually good to communicate with all stakeholders earlier rather than later about the assumption [...]."} 
While it is not required to document all the information at the very early stages of development, our recommendation can be done in the timely manner.
In other words, before iteration planning, practitioners can check whether these useful types of information are documented, correct, and up to date to avoid making unnecessary assumptions.

\textbf{Future research should focus on developing tools and techniques to assess the quality of \di{} for effort estimation.}
While it is recommended to maintain the documentation for effort estimation, it may still be tedious for practitioners to assess the quality of information of all tasks.
Techniques to detect missing or changing information should help practitioners to timely manage the useful \di{}, which in turn will lead to more accurate effort estimation.
Yet, less focus has been given to developing techniques that detect missing or changing information that is useful for effort estimation.
To the best of our knowledge, existing work focused on developing techniques to detect missing information for software development, e.g., the information for bug-fixing tasks~\cite{Chaparro2017}, feature requests in issue tracking systems~\cite{merten2016software}, and user stories from developer-client conversation transcripts~\cite{rodeghero2017detecting}.
Therefore, future research should develop techniques that help practitioners be aware of the quality issues of \di{} that could impact the accuracy of effort estimation.

    \section{Related Work}\label{sec:relatedwork}

In this section, we discuss related work with respect to effort estimation and documentation.

\textbf{Information and Estimation.} Several studies investigated the impact of inadequate information on effort estimation.
Jørgensen et al.~\cite{jorgensen2009impact} found that unclear task information was one of the reasons for inaccurate estimation.
Svensson et al.~\cite{svensson2019bam} pointed out that inadequate information could lead the team to misunderstood the task, which may lead to inaccurate estimation.
Usman et al.~\cite{Usman2016, usman2018effort} reported that Agile practitioners perceived that a lack of details in requirements could lead to inaccurate estimation.
Hoda and Murugesan~\cite{hoda2016multilevel} conducted a grounded theory study with Agile practitioners and reported that some information could be missing or delayed, which could lead to a wrong assumption and inaccurate estimation.
Bick et al.~\cite{Bick2018} conducted a case study with Agile teams and reported that a task is re-estimated when the information change caused the estimated effort to became unrealistic.
Complimenting the prior studies, this work shows that a change of \di{} is more likely to influence re-estimation than a change of \ios{}.

\textbf{Documentation in Agile.}
Documentation is one of the approaches to record and convey information related to a software product, which a team can use as a reference~\cite{andriyani2017understanding}.
Sedano et al.~\cite{sedano2019product} reported that a product backlog could be used to remind the work to be done and assist the face-to-face communication.
Kasauli et al.~\cite{kasauli2017requirements} reported that practitioners relied on information recorded in documentation (i.e., requirements specifications, backlog items) to build and maintain understanding for several purposes, e.g., task implementation, system maintenance, defining test cases.
\revisedTextOnly{Aligning with the prior work, our results highlight that documentation is commonly used for effort estimation and \di{} is considered moderately to extremely important when estimating effort.}

\textbf{Minimal Documentation.}
Comprehensive documentation is considered overhead based on the Agile manifesto~\cite{fowler2001agile}.
On the other hand, Kasauli et al.~\cite{kasauli2017requirements} discussed that information might be inadequate for developers if the team aimed to minimize documentation effort by only documenting \dius{} and test cases.
Ernst and Murphy~\cite{ernst2012case} reported that information could be documented in a just-in-time fashion, i.e., requirements were first sketched out with simple natural language statements and then fully elaborated just before or during development.
Heck and Zaidman~\cite{heck2017framework} found that some types of information (e.g., use cases, UI mock-ups) might not be mandatory at the early stage.
Motivated by the prior studies, we identified the useful types of information that should be documented for effort estimation, i.e., \difr{}, \dius{}, \didd{}, \diui{}, \diat{}, and \ditd{}.

\textbf{Quality Issues.} Several studies analyzed the quality of documentation.
Aghajani et al.~\cite{aghajani2019software} manually analyzed 878 documentation-related artifacts and found that \di{} was often found incorrect, incomplete, and outdated.
Kasauli et al.~\cite{kasauli2017requirements} reported that the information documented in backlog items was often changing while the information documented in requirement specifications was often outdated.
Prior studies also reported that \di{} could be missing or did not match the developer needs for bug-fixing tasks~\cite{Chaparro2017, Zimmermann2010}.
Similarly, in the context of Agile effort estimation, we found that the useful types of \di{} for effort estimation were occasionally (or more) \qchange{} or \qmiss{} during or after effort estimation.

\section{Threats to Validity}\label{sec:threats}

\textbf{Construct validity} is related to the concerns on the survey design. 
The survey respondents might misinterpret the survey questions or definitions of the terms.
To mitigate this threat, we conducted two pilot tests with five experienced software engineers to check the comprehensibility of our survey questions.
We also provided the definitions of terms and the nine types of \di{} we used in the survey.
In addition, we informed the invitees that they could directly contact the authors for clarification or requesting support via video calls.

It might also be possible that the term "\difr{}" might be misinterpreted as the software requirement specification (SRS) of the non-agile processes.
However, our study focused only on the software engineers who used Agile methods and we asked them to answer the survey questions at the lowest granularity that they estimate.
We also confirmed the interpretation with the pilot test participants and they explained that the \difr{} were understood as \frdefinition{}.
Hence, it is less likely that the \difr{} were interpreted as the SRS.

\textbf{Internal validity} concerns the confounding factors that might impact our findings.
Our survey results might be varied based on other factors, e.g., company practices or product domain.
However, we did not ask the participants about their company or product because such sensitive information might cause them to be reluctant to participate, increased the risk of evaluation apprehension~\cite{ghazi2018survey} and self-selection bias~\cite{kasunic2005designing}.
Some invitees also responded that they would not participate in the survey if it collects such information.
Other study methods (e.g., interviews, Delphi sessions) might provide additional insights about the variation based on other factors.

\revised{R1.7}{
In RQ2, the responses of Q5-8 might be biased toward re-estimation if \di{} was changed since most of our respondents consider documentation important (RQ1).
To check this, we used the Spearman rank correlation test to measure whether the responses of Q2 and Q5-8 are highly correlated (i.e., the higher the perceived importance of \di{}, the more likely the re-estimation is performed when \di{} was changed).
We found that the correlations are small to moderate (i.e., $|\rho|$ is 0.2-0.39).
Hence, the responses in RQ2 are less likely biased.
}


The first two authors conducted the analysis of open-ended answers (i.e., card sorting).
Personal experiences of the authors might influence the analysis.
Therefore, we validated the analysis by the closed-card sorting performed by the third author.
The Cohen's Kappa coefficient~\cite{mchugh2012interrater} ranged from 0.65-0.80, indicating substantial agreements between the sorters.

When we performed statistical tests, we converted the Likert scale into numerical values.
The statistical test results could be impacted because the distance between the scale values were not ordinal.
However, as suggested by Harpe~\cite{jamieson2004likert} and Jamieson~\cite{harpe2015analyze}, it is a common practice to treat rating items with at least five numerical categories as continuous data.

\textbf{External validity} is related to the concern on the generalizability of our findings.
Despite the diverse background of our respondents (c.f. Section~\ref{sec:demographic}), the survey results might not generalize to all kinds of Agile projects or effort estimation practices.
Nevertheless, to facilitate a replicate study of this work, we provided a full set of survey questions and relevant documents in the replication package.\textsuperscript{\ref{drive}}
    \section{Conclusions}\label{sec:conclusions}

Effort estimation is an important practice of Agile planning activities.
To estimate the effort, practitioners use the information to understand the task.
One of the common approaches to convey the information is to use documentation.
However, documentation has a lower priority than delivering working software in the Agile practices. 
It is possible that documentation may not be well maintained and may impact the accuracy of effort estimation.
Yet, little is known about how documented effort can be minimized while achieving accurate estimation.
Hence, in this paper, we investigated the importance of documentation and the types of \di{} that are useful for effort estimation in the current Agile practices.
Through a survey study with \eligibleCount{} Agile practitioners, we found that:

\begin{itemize}
    \item Despite the low priority of documentation in Agile, a majority of our respondents used documentation to assist effort estimation and considered the \di{} as moderately to extremely important when estimating effort.
    \item The reported frequency of re-estimation when the \di{} was changed is statistically higher than the reported frequency of re-estimation when \ios{} was changed. Furthermore, their re-estimation decision was based on the significance or the source of the \infochg{}. 
    \item Functional requirements, \dius{}, \didd{}, \diui{}, \diat{}, and \ditd{} were ranked as the six most useful \di{} for effort estimation, however many respondents reported that these useful types of information were occasionally (or more) \qchange{} or \qmiss{} during (or after) effort estimation.
\end{itemize}

The key contribution of our work is to shed light on the importance of \di{} and its impact on the accuracy of effort estimation in Agile practices.
Our findings suggest that the quality of \di{} can have an impact on effort estimation.
Nevertheless, we do not expect Agile practitioners to prepare comprehensive documentation.
Indeed, to achieve a practice of minimal documentation of Agile, our work highlights the six types of \di{} that practitioners should pay attention before performing effort estimation. 
Our recommendations can be done in a timely manner, which helps practitioners achieve more accurate effort estimation with an optimal maintenance effort of documentation.

    \newline
    \textbf{Acknowledgement.} P.  Thongtanunam  was  partially  supported  by  the  Australian
Research Council’s Discovery Early Career Researcher Award
(DECRA) funding scheme (DE210101091).
    \bibliographystyle{IEEEtranS}
    \bibliography{main}

\begin{thebibliography}{10}
\providecommand{\url}[1]{#1}
\csname url@samestyle\endcsname
\providecommand{\newblock}{\relax}
\providecommand{\bibinfo}[2]{#2}
\providecommand{\BIBentrySTDinterwordspacing}{\spaceskip=0pt\relax}
\providecommand{\BIBentryALTinterwordstretchfactor}{4}
\providecommand{\BIBentryALTinterwordspacing}{\spaceskip=\fontdimen2\font plus
\BIBentryALTinterwordstretchfactor\fontdimen3\font minus
  \fontdimen4\font\relax}
\providecommand{\BIBforeignlanguage}[2]{{%
\expandafter\ifx\csname l@#1\endcsname\relax
\typeout{** WARNING: IEEEtranS.bst: No hyphenation pattern has been}%
\typeout{** loaded for the language `#1'. Using the pattern for}%
\typeout{** the default language instead.}%
\else
\language=\csname l@#1\endcsname
\fi
#2}}
\providecommand{\BIBdecl}{\relax}
\BIBdecl

\bibitem{IEEETerm}
``{ISO/IEC/IEEE International Standard - Systems and software engineering --
  Vocabulary},'' \emph{ISO/IEC/IEEE 24765:2010(E)}, pp. 1--418, 2010.

\bibitem{aghajani2019software}
E.~Aghajani, C.~Nagy, O.~L. Vega-M{\'a}rquez, M.~Linares-V{\'a}squez,
  L.~Moreno, G.~Bavota, and M.~Lanza, ``{Software Documentation Issues
  Unveiled},'' in \emph{Proc. of the ICSE}, 2019, pp. 1199--1210.

\bibitem{andriyani2017understanding}
Y.~Andriyani, R.~Hoda, and R.~Amor, ``{Understanding Knowledge Management in
  Agile Software Development Practice},'' in \emph{Proc. of the KSEM}, 2017,
  pp. 195--207.

\bibitem{behutiye2017non}
W.~Behutiye, P.~Karhap{\"a}{\"a}, D.~Costal, M.~Oivo, and X.~Franch,
  ``{Non-functional Requirements Documentation in Agile Software Development:
  Challenges and Solution Proposal},'' in \emph{Proc. of the PROFES}, 2017, pp.
  515--522.

\bibitem{Bick2018}
S.~{Bick}, K.~{Spohrer}, R.~{Hoda}, A.~{Scheerer}, and A.~{Heinzl},
  ``{Coordination Challenges in Large-Scale Software Development: A Case Study
  of Planning Misalignment in Hybrid Settings},'' \emph{TSE}, vol.~44, no.~10,
  pp. 932--950, 2018.

\bibitem{Chaparro2017}
O.~Chaparro, J.~Lu, F.~Zampetti, L.~Moreno, M.~Di~Penta, A.~Marcus, G.~Bavota,
  and V.~Ng, ``{Detecting Missing Information in Bug Descriptions},'' in
  \emph{Proc. of the ESEC/FSE}, 2017, pp. 396--407.

\bibitem{coelho2012effort}
E.~Coelho and A.~Basu, ``{Effort Estimation in Agile Software Development using
  Story Points},'' \emph{IJAIS}, vol.~3, no.~7, pp. 7--10, 2012.

\bibitem{Cohn2006}
M.~Cohn, \emph{{Agile estimating and planning}}.\hskip 1em plus 0.5em minus
  0.4em\relax {Pearson Education}, 2006.

\bibitem{Davies2014}
S.~Davies and M.~Roper, ``{What's in a Bug Report?}'' in \emph{{Proc. of the
  ESEM}}, no.~26, 2014, pp. 1--10.

\bibitem{denscombe2014good}
M.~Denscombe, \emph{The Good Research Guide: for small-scale social research
  projects}.\hskip 1em plus 0.5em minus 0.4em\relax McGraw-Hill Education (UK),
  2014.

\bibitem{ebert2008effectively}
C.~Ebert and J.~De~Man, ``{Effectively utilizing project, product and process
  knowledge},'' \emph{IST}, vol.~50, no.~6, pp. 579--594, 2008.

\bibitem{ebert2019confusion}
F.~Ebert, F.~Castor, N.~Novielli, and A.~Serebrenik, ``{Confusion in code
  reviews: Reasons, impacts, and coping strategies},'' in \emph{Proc. of the
  SANER}, 2019, pp. 49--60.

\bibitem{ernst2012case}
N.~A. Ernst and G.~C. Murphy, ``{Case Studies in Just-In-Time Requirements
  Analysis},'' in \emph{Proc. of the EmpiRE}, 2012, pp. 25--32.

\bibitem{etikan2016comparison}
I.~Etikan, S.~A. Musa, and R.~S. Alkassim, ``{Comparison of Convenience
  Sampling and Purposive Sampling},'' \emph{AJTAS}, vol.~5, no.~1, pp. 1--4,
  2016.

\bibitem{forward2002relevance}
A.~Forward and T.~C. Lethbridge, ``{The Relevance of Software Documentation,
  Tools and Technologies: A Survey},'' in \emph{Proc. of the ACM Symposium on
  Document Engineering}, 2002, pp. 26--33.

\bibitem{fowler2001agile}
M.~Fowler and J.~Highsmith, ``The agile manifesto,'' \emph{Software
  Development}, vol.~9, no.~8, pp. 28--32, 2001.

\bibitem{fucci2018needs}
D.~Fucci, C.~Palomares, X.~Franch, D.~Costal, M.~Raatikainen, M.~Stettinger,
  Z.~Kurtanovic, T.~Kojo, L.~Koenig, A.~Falkner \emph{et~al.}, ``{Needs and
  Challenges for a Platform to Support Large-scale Requirements Engineering: a
  Multiple-case Study},'' in \emph{Proc. of the ESEM}, no.~19, 2018, pp. 1--10.

\bibitem{ghazi2018survey}
A.~N. Ghazi, K.~Petersen, S.~S. V.~R. Reddy, and H.~Nekkanti, ``{Survey
  Research in Software Engineering: Problems and Mitigation Strategies},''
  \emph{IEEE Access}, vol.~7, pp. 1--24, 2018.

\bibitem{gralha2018evolution}
C.~Gralha, D.~Damian, A.~Wasserman, M.~Goul{\~a}o, and J.~Ara{\'u}jo, ``{The
  Evolution of Requirements Practices in Software Startups},'' in \emph{Proc.
  of the ICSE}, 2018, pp. 823--833.

\bibitem{harpe2015analyze}
S.~E. Harpe, ``{How to analyze Likert and other rating scale data},''
  \emph{CPTL}, vol.~7, no.~6, pp. 836--850, 2015.

\bibitem{heck2017framework}
P.~Heck and A.~Zaidman, ``{A framework for quality assessment of just-in-time
  requirements: the case of open source feature requests},'' \emph{RE},
  vol.~22, no.~4, pp. 453--473, 2017.

\bibitem{heck2018systematic}
------, ``{A Systematic Literature Review on Quality Criteria for Agile
  Requirements Specifications},'' \emph{SQJ}, vol.~26, no.~1, pp. 127--160,
  2018.

\bibitem{hoda2016multilevel}
R.~Hoda and L.~K. Murugesan, ``{Multi-Level Agile Project Management
  Challenges: A Self-Organizing Team Perspective},'' \emph{JSS}, vol. 117, pp.
  245--257, 2016.

\bibitem{hoda2012documentation}
R.~Hoda, J.~Noble, and S.~Marshall, ``{Documentation Strategies on Agile
  Software Development Projects},'' \emph{IJAESD}, vol.~1, no.~1, pp. 23--37,
  2012.

\bibitem{inayat2015systematic}
I.~Inayat, S.~S. Salim, S.~Marczak, M.~Daneva, and S.~Shamshirband, ``{A
  systematic literature review on agile requirements engineering practices and
  challenges},'' \emph{Computers in human behavior}, vol.~51, pp. 915--929,
  2015.

\bibitem{james2010scrum}
M.~James and L.~Walter, ``{Scrum Reference Card},'' \emph{CollabNet}, 2010.

\bibitem{jamieson2004likert}
S.~Jamieson, ``{Likert scales: How to (ab) use them?}'' \emph{Medical
  education}, vol.~38, no.~12, pp. 1217--1218, 2004.

\bibitem{jorgensen2009impact}
M.~J{\o}rgensen and T.~M. Gruschke, ``{The Impact of Lessons-Learned Sessions
  on Effort Estimation and Uncertainty Assessments},'' \emph{TSE}, vol.~35,
  no.~3, pp. 368--383, 2009.

\bibitem{kasauli2017requirements}
R.~Kasauli, G.~Liebel, E.~Knauss, S.~Gopakumar, and B.~Kanagwa, ``{Requirements
  Engineering Challenges in Large-Scale Agile System Development},'' in
  \emph{RE}, 2017, pp. 352--361.

\bibitem{kasunic2005designing}
M.~Kasunic, ``{Designing an effective survey},'' Carnegie-Mellon Univ
  Pittsburgh PA Software Engineering Inst, Tech. Rep., 2005.

\bibitem{khalil2019exploring}
C.~Khalil and S.~Khalil, ``{Exploring knowledge management in agile software
  development organizations},'' \emph{IEMJ}, vol.~16, no.~2, pp. 555--569,
  2020.

\bibitem{kitchenham2002preliminary}
B.~A. Kitchenham, S.~L. Pfleeger, L.~M. Pickard, P.~W. Jones, D.~C. Hoaglin,
  K.~El~Emam, and J.~Rosenberg, ``{Preliminary Guidelines for Empirical
  Research in Software Engineering},'' \emph{TSE}, vol.~28, no.~8, pp.
  721--734, 2002.

\bibitem{klotins2019progression}
E.~Klotins, M.~Unterkalmsteiner, P.~Chatzipetrou, T.~Gorschek, R.~Prikladnicki,
  N.~Tripathi, and L.~Pompermaier, ``{A progression model of software
  engineering goals, challenges, and practices in start-ups},'' \emph{TSE}, p.
  To appear, 2019.

\bibitem{lauesen2002software}
S.~Lauesen, \emph{{Software Requirements: Styles and Techniques}}.\hskip 1em
  plus 0.5em minus 0.4em\relax Pearson Education, 2002.

\bibitem{mchugh2012interrater}
M.~L. McHugh, ``{Interrater reliability: the kappa statistic},''
  \emph{Biochemia medica: Biochemia medica}, vol.~22, no.~3, pp. 276--282,
  2012.

\bibitem{merten2016software}
T.~Merten, M.~Falis, P.~H{\"u}bner, T.~Quirchmayr, S.~B{\"u}rsner, and
  B.~Paech, ``{Software Feature Request Detection in Issue Tracking Systems},''
  in \emph{Proc. of the RE}, 2016, pp. 166--175.

\bibitem{paetsch2003requirements}
F.~Paetsch, A.~Eberlein, and F.~Maurer, ``{Requirements Engineering and Agile
  Software Development},'' in \emph{Proc. of the WETICE}, 2003, pp. 308--313.

\bibitem{palomares2017requirements}
C.~Palomares, C.~Quer, and X.~Franch, ``{Requirements Reuse and Requirement
  Patterns: A State of the Practice Survey},'' \emph{ESE}, vol.~22, no.~6, pp.
  2719--2762, 2017.

\bibitem{rodeghero2017detecting}
P.~Rodeghero, S.~Jiang, A.~Armaly, and C.~McMillan, ``{Detecting User Story
  Information in Developer-Client Conversations to Generate Extractive
  Summaries},'' in \emph{Proc. of the ICSE}, 2017, pp. 49--59.

\bibitem{rubin2012essential}
K.~S. Rubin, \emph{{Essential Scrum: A Practical Guide to the Most Popular
  Agile Process}}.\hskip 1em plus 0.5em minus 0.4em\relax Addison-Wesley, 2012.

\bibitem{schon2017agile}
E.-M. Sch{\"o}n, J.~Thomaschewski, and M.~J. Escalona, ``{Agile Requirements
  Engineering: A systematic literature review},'' \emph{Computer Standards \&
  Interfaces}, vol.~49, pp. 79--91, 2017.

\bibitem{sedano2019product}
T.~Sedano, P.~Ralph, and C.~P{\'e}raire, ``{The Product Backlog},'' in
  \emph{Proc. of the ICSE}, 2019, pp. 200--211.

\bibitem{spencer2009card}
D.~Spencer, \emph{{Card sorting: Designing usable categories}}.\hskip 1em plus
  0.5em minus 0.4em\relax Rosenfeld Media, 2009.

\bibitem{svensson2019bam}
R.~B. Svensson, T.~Gorschek, P.-O. Bengtsson, and J.~Widerberg, ``{BAM-Backlog
  Assessment Method},'' in \emph{Proc. of the XP}, 2019, pp. 53--68.

\bibitem{tanveer2016understanding}
B.~Tanveer, L.~Guzm{\'a}n, and U.~M. Engel, ``{Understanding and improving
  effort estimation in agile software development: an industrial case study},''
  in \emph{Proc. of the ICSSP}, 2016, pp. 41--50.

\bibitem{tizard2019can}
J.~Tizard, H.~Wang, L.~Yohannes, and K.~Blincoe, ``{Can a Conversation Paint a
  Picture? Mining Requirements in Software Forums},'' in \emph{Proc. of the
  RE}, 2019, pp. 17--27.

\bibitem{Usman2016}
M.~{Usman} and R.~{Britto}, ``{Effort Estimation in Co-located and Globally
  Distributed Agile Software Development: A Comparative Study},'' in
  \emph{Proc. of the IWSM-MENSURA}, 2016, pp. 219--224.

\bibitem{usman2018effort}
M.~Usman, R.~Britto, L.-O. Damm, and J.~B{\"o}rstler, ``Effort estimation in
  large-scale software development: An industrial case study,'' \emph{IST},
  vol.~99, pp. 21--40, 2018.

\bibitem{vassallo2020developers}
C.~Vassallo, S.~Panichella, F.~Palomba, S.~Proksch, H.~C. Gall, and A.~Zaidman,
  ``{How developers engage with static analysis tools in different contexts},''
  \emph{{EMSE}}, vol.~25, no.~2, pp. 1419--1457, 2020.

\bibitem{wang2014role}
X.~Wang, L.~Zhao, Y.~Wang, and J.~Sun, ``{The Role of Requirements Engineering
  Practices in Agile Development: An Empirical Study},'' \emph{RE}, vol. 432,
  pp. 195--209, 2014.

\bibitem{xia2019practitioners}
X.~Xia, Z.~Wan, P.~S. Kochhar, and D.~Lo, ``{How Practitioners Perceive Coding
  Proficiency},'' in \emph{Proc. of the ICSE}, 2019, pp. 924--935.

\bibitem{zhi2015cost}
J.~Zhi, V.~Garousi-Yusifo{\u{g}}lu, B.~Sun, G.~Garousi, S.~Shahnewaz, and
  G.~Ruhe, ``Cost, benefits and quality of software development documentation:
  A systematic mapping,'' \emph{JSS}, vol.~99, pp. 175--198, 2015.

\bibitem{Zimmermann2010}
T.~Zimmermann, R.~Premraj, N.~Bettenburg, S.~Just, A.~Schroter, and C.~Weiss,
  ``{What Makes a Good Bug Report?}'' \emph{TSE}, vol.~36, no.~5, pp. 618--643,
  2010.

\end{thebibliography}

\end{document}